\documentclass[paper]{JHEP}
\usepackage{epsfig}
\usepackage{amssymb}
\def\beq{\begin{equation}}
\def\eeq{\end{equation}}

\def\beqn{\begin{eqnarray}}
\def\eeqn{\end{eqnarray}}

\def\HW{{\small HERWIG}}
\def\NLO{{\small NLO}}
\def\MC{{\small MC}}

\newcommand\sss{\scriptscriptstyle\rm}

\newcommand\MCatNLO{{\rm MC}@{\rm NLO}}

\newcommand\xMC{|_{\sss {\rm MC}}}

\newcommand\bSigma{\overline{\Sigma}}

\newcommand\code{\tt}
\newcommand\variable{\tt}
\preprint{
 Cavendish--HEP--02/09\hfill\\
 LAPTH--922/02\hfill\\
 GEF-th-09/2002\hfill\\
 hep-ph/0207182}
\title{\boldmath The MC@NLO Event Generator%
\footnote{Work supported in part by the UK Particle Physics and
Astronomy Research Council and by the EU Fourth Framework Programme
`Training and Mobility of Researchers', Network `Quantum Chromodynamics
and the Deep Structure of Elementary Particles',
contract FMRX-CT98-0194 (DG 12 - MIHT).}}
\author{Stefano Frixione%
  \thanks{On leave of absence from INFN, Sez. di Genova, Italy}\\
  Laboratoire d'Annecy-le-Vieux de Physique de Particules\\
  Chemin de Bellevue, BP 110,
  74941 Annecy-le-Vieux CEDEX, France\\
  E-mail: \email{Stefano.Frixione@cern.ch}}
\author{Bryan R.\ Webber\\
  Cavendish Laboratory, University of Cambridge,\\
  Madingley Road, Cambridge CB3 0HE, U.K.\\
  E-mail: \email{webber@hep.phy.cam.ac.uk}}
\abstract{
This is the user's manual of {$\MCatNLO$}.1.0. This package is a 
practical implementation, based upon the HERWIG event generator,
of the recently proposed $\MCatNLO$ formalism for matching the
next-to-leading order calculation of a QCD process with a
parton-shower Monte Carlo simulation. The processes of 
standard-model vector boson pair production in hadronic
collisions are available.
}
\begin{document}

\section{Generalities}
In this documentation file, we briefly describe how to run the 
$\MCatNLO$, implemented according to the formalism introduced in 
ref.~\cite{Frixione:2002ik}. The following production processes 
are now available ({\variable IPROC} has the same meaning as in 
\HW\ \cite{Marchesini:1992ch}):
\begin{table}[htb]
\begin{center}
\begin{tabular}{|c|l|}\hline
{\variable IPROC} & Process \\\hline
 2850 & $H_1 H_2\to$ W$^+$W$^-+X$\\\hline
 2860 & $H_1 H_2\to$ ZZ$+X$\\\hline
 2870 & $H_1 H_2\to$ W$^+$Z$+X$\\\hline
 2880 & $H_1 H_2\to$ W$^-$Z$+X$\\\hline
\end{tabular}
\end{center}
\caption{\label{tab:proc}
Processes implemented in $\MCatNLO$.
}
\end{table}

\subsection{Mode of operation}
In the case of standard MC, a hard kinematical configuration is
generated on a event-by-event basis, and it is subsequently showered 
and hadronized. In the case of our $\MCatNLO$, all of the hard kinematical
configurations are generated in advance, and stored in a file 
(which we call {\em event file} -- see sect.~\ref{sec:evfile}); 
the event file is then read by \HW, which showers and hadronizes each 
hard configuration. Therefore, in the $\MCatNLO$ the reading of a 
hard configuration from the event file is equivalent to the generation 
of such a configuration in the standard MC. Apart from this
difference, $\MCatNLO$ and MC behave exactly in the same way. Thus,
the available user's analysis routines can be used without any
modification in the case of $\MCatNLO$. One should recall, however,
that $\MCatNLO$ always generates some events with negative weights
(see ref.~\cite{Frixione:2002ik}); therefore, the correct distributions
are obtained by summing weights with their signs (i.e., the absolute 
values of the weights must {\em NOT} be used when filling the histograms).

With such a structure, it is natural to create two separate executables,
which we improperly denote as \NLO\ and \MC. The former has the sole scope
of creating the event file; the latter is just \HW, augmented by
the capability of reading the event file.

\subsection{Package files}
The package consists of the following files:

\begin{itemize}
\item {\bf Shell utilities}\\
    {\code MCatNLO.Script}\\
    {\code MCatNLO.inputs}\\
    {\code Makefile}

\item {\bf General purpose codes}\\
    {\code alpha.f}\\ 
    {\code dummies.f}\\ 
    {\code linux.f}\\ 
    {\code mcatnlo\_date.f}\\  
    {\code mcatnlo\_int.f}\\
    {\code mcatnlo\_libofpdf.f}\\ 
    {\code mcatnlo\_mlmtopdf.f}\\ 
    {\code mcatnlo\_pdftomlm.f}\\ 
    {\code mcatnlo\_str.f}\\ 
    {\code mcatnlo\_uti.f}\\ 
    {\code mcatnlo\_uxdate.c}\\
    {\code sun.f}\\ 
    {\code trapfpe.c}

\item {\bf Vector boson pair production codes}\\
    {\code mcatnlo\_hwanal.f}\\
    {\code mcatnlo\_hwdriver.f}\\ 
    {\code mcatnlo\_hwhvvj.f}\\ 
    {\code mcatnlo\_vbmain.f}\\
    {\code mcatnlo\_vbxsec.f}
\end{itemize}
These files can be downloaded from the Web page:\\
$\phantom{aaaaaaaa}$%
{\code http://www.hep.phy.cam.ac.uk/theory/webber/MCatNLO}\\
The vector boson pair production codes and shell utilities are only
relevant to the production of W$^+$W$^-$, ZZ, and W$^\pm$Z pairs in hadronic
collisions. The general purpose codes will be used in the $\MCatNLO$
implementation of other production processes.

In addition to the files listed above, the user will need a version of
the \HW\ code, which can be downloaded from the Web page:\\
$\phantom{aaaaaaaa}$%
{\code http://hepwww.rl.ac.uk/theory/seymour/herwig/}\\
As stressed in 
ref.~\cite{Frixione:2002ik}, for the $\MCatNLO$ we don't need to 
modify the existing (LL) shower algorithm; thus, users can simply link 
(some of) the files above to their preferred \HW\ version.  To be
capable of handling the relevant event process codes listed in
table~\ref{tab:proc}, this should be public version 6.301 or higher.
On most systems, users will need to delete the dummy subroutines
{\small HWHVVJ}, {\small PDFSET} and {\small STRUCTM} from the
standard \HW\ package, to permit linkage of the corresponding routines
from the $\MCatNLO$ package. 

\subsection{Working environment}
We have written a number of shell scripts and a {\code Makefile} (all
listed under {\bf Shell utilities} above) which will simplify the use of
the package considerably. To use them, the computing system
must support {\code bash} shell, and {\code gmake}. 
Should these be unavailable on
the user's computing system, the compilation and running of our
$\MCatNLO$ requires more detailed instructions, for which we
refer the reader to app.~\ref{app:instr}. This appendix will also
serve as a reference for more advanced usage of the package.

\subsection{Source and running directories}
We assume that all the files of the package sit in the same directory,
which we call the {\em source directory}. When creating the executable, 
our shell scripts determine the type of operating system, and create a
subdirectory of the source directory, which we call the {\em running 
directory}, whose name is {\variable Alpha}, {\variable Sun}, or {\variable
Linux}, depending on the operating system.  If the operating system is
not known by our scripts, the name of the working directory is
{\variable Run}. The running directory contains all the object files
and executable files, and in general all the files produced by the
$\MCatNLO$ while running.  It must also contain the relevant grid
files (see sect.~\ref{sec:pdfs}), or links to them, if the library of
parton densities provided in the package is used. 

\section{Prior to running}
Before running the code, the user needs to edit the following files:\\
$\phantom{aaa}${\code mcatnlo\_hwanal.f}\\
$\phantom{aaa}${\code mcatnlo\_hwdriver.f}\\ 
$\phantom{aaa}${\code mcatnlo\_hwhvvj.f}\\
$\phantom{aaa}${\code Makefile}\\
We do not assume that the user will adopt the latest release of 
\HW\ (although we do recommend such a choice). For this reason,
the files {\code mcatnlo\_hwdriver.f} and {\code mcatnlo\_hwhvvj.f} 
must be edited, in order to modify the 'INCLUDE HERWIGXX.INC' command
to correspond to the version of \HW\ the user is going to adopt.
{\code mcatnlo\_hwdriver.f} contains a set of read statements,
which are necessary for the \MC\ to get the input parameters (see
sect.~\ref{sec:running} for the input procedure); these read
statements must not be modified or eliminated. Also, {\code
mcatnlo\_hwdriver.f} calls the \HW\ routines which
perform showering, hadronization, decays, and so forth; the user can
freely modify this part, as customary in MC runs. Finally, the sample
code {\code mcatnlo\_hwanal.f} contains analysis-related routines:
this file must be replaced by a file which contains the user's analysis 
routines.

The {\code Makefile} must also be edited, in order to set two variables,
{\variable HERWIGVER} and {\variable HWUTI}. The former variable must be set 
equal to the object file name of the version of \HW\ currently adopted 
(matching the one whose common blocks are included in the files
above). The variable {\variable HWUTI} must be set equal to the list
of object files that the user needs in the analysis routines. The
lists of files linked by {\code Makefile} are reported in
app.~\ref{app:exe}. See also sect.~\ref{sec:Makevar}.

\subsection{Parton densities\label{sec:pdfs}}
Since the knowledge of the parton densities (PDFs) is necessary in
order to get the physical cross section, a PDF library must be
linked. The possibility exists to link the CERNLIB PDF library
(PDFLIB); however, we also provide a self-contained PDF library with
this package, which is faster than PDFLIB. The user may link either
PDF library; all that is necessary is to set the variable {\variable
PDFLIBRARY} (in the file {\code MCatNLO.inputs}) equal to {\variable
THISLIB} if one wants to link to our PDF library, and equal to
{\variable PDFLIB} if one wants to link to PDFLIB.  Our PDF library
collects the original codes, written by the authors of the PDF fits;
as such, for most of the densities it needs to read the files which
contain the grids that initialize the PDFs. These files, also provided
with the present package, must either be copied into the running
directory, or defined in the running directory as logical
links to the physical files (by using {\code ln -sn}).

As stressed before, consistent inputs must be given to the \NLO\ and
\MC\ codes. However, in ref.~\cite{Frixione:2002ik} we found that the
dependence upon the PDFs used by the MC is rather weak. So one may
want to run the \NLO\ and \MC\ adopting a regular NLO-evolved set in the
former case, and the default (LO) \HW\ set in the latter (the advantage
is that this option reduces the amount of running time of the MC). In
order to do so, the user must set the variable {\variable HERPDF}
equal to {\variable DEFAULT} in the file {\code MCatNLO.inputs};
setting {\variable HERPDF=EXTPDF} will force the MC to use the same
PDF set as the NLO code.

Regardless of the PDFs used in the MC run, users must delete the dummy 
PDFLIB routines {\small PDFSET} and {\small STRUCTM} from \HW, as
explained earlier.

\section{Running\label{sec:running}}
It is straightforward to run the $\MCatNLO$. First, edit\\
$\phantom{aaa}${\code MCatNLO.inputs}\\
and write there all the input parameters (for the complete list 
of the input parameters, see sect.~\ref{sec:scrvar}). As the last
line of the file {\code MCatNLO.inputs}, write\\
$\phantom{aaa}${\code runMCatNLO}\\
Finally, execute {\code MCatNLO.inputs} from the (bash) shell.
This procedure will create the NLO and MC executables, and run them
using the inputs given in {\code MCatNLO.inputs}, which guarantees
that the parameters used in the \NLO\ and \MC\ runs are identical.
Should the user only need to create the executables without running
them, or to run the \NLO\ or \MC\ only, he/she should replace the
call to {\code runMCatNLO} in the last line of {\code MCatNLO.inputs}
by calls to\\
$\phantom{aaa}${\code compileNLO}\\
$\phantom{aaa}${\code compileMC}\\
$\phantom{aaa}${\code runNLO}\\
$\phantom{aaa}${\code runMC}\\
which have obvious meanings ({\code runXX} also creates the {\code XX}
executable).

We stress that the input parameters are not solely related to
physics (masses, CM energy, and so on); there are a few of them
which control other things, such as the number of events generated.
These must also be set by the user, according to his/her needs:
see sect.~\ref{sec:scrvar}.

If the shell scripts are not used to run the codes, the inputs are
given to the \NLO\ or \MC\ codes during an interactive talk-to phase;
the complete sets of inputs for our codes are reported in 
app.~\ref{app:input}.

\subsection{Event file\label{sec:evfile}}
The NLO code creates the event file. In order to do so, it goes through
two steps; first it integrates the cross sections (the {\em integration
step}), and then, using the information gathered in the integration step, 
produces a set of events (the {\em event generation step}).

The event generation step necessarily follows the integration step;
however, for each integration step one can have an arbitrary number of
event generation steps, i.e., an arbitrary number of event files.
This is useful in the case in which the statistics accumulated 
with a given event file is not sufficient.

Suppose the user wants to create an event file; editing {\code
MCatNLO.inputs}, the user sets {\variable BASES=ON}, to enable the
integration step, sets the parameter {\variable NEVENTS} equal to
the number of events wanted on tape, and runs the code; the
information on the integration step (unreadable to the user, but
needed by the code in the event generation step) is written on files
whose name begin with {\variable FPREFIX}, a string the user sets
in {\code MCatNLO.inputs}; these files (which we denote as {\em data
files}) have extensions {\code .data}. The name of the event file is 
{\variable EVPREFIX.events}, where {\variable EVPREFIX} is again a 
string set by the user.

Now suppose the user wants to create another event file, to increase
the statistics. The user simply sets {\variable BASES=OFF}, since 
the integration step is not necessary any longer (however, the data
files must not be removed: the information
stored there is still used by the NLO code); changes the string
{\variable EVPREFIX} (failure to do so overwrites the existing event
file), while keeping {\variable FPREFIX} at the same value as before;
and changes the value of {\variable RNDEVSEED} (the random number
seed used in the event generation step; failure to do so results in
an event file identical to the previous one); the number {\variable
NEVENTS} generated may or may not be equal to the one chosen in
generating the former event file(s).

We point out that data and event files may be very large. If the user
wants to store them in a scratch area, this can be done by setting the
script variable {\variable SCRTCH} equal to the physical address
of the scratch area (see sect.~\ref{sec:res}).

\subsection{Results\label{sec:res}}
As in the case of standard \HW, the form of the results will be
determined by the user's analysis routines. However, in addition
to any files written by the user's analysis routines, the
$\MCatNLO$ writes the following files:\\
$\blacklozenge$ 
{\variable FPREFIXNLOinput}: the input file for the \NLO\ executable, 
created according to the set of input parameters defined in 
{\code MCatNLO.inputs} (where the user also sets the string
{\variable FPREFIX}). See table~\ref{tab:NLOi}.\\
$\blacklozenge$ 
{\variable FPREFIXNLO.log}: the log file relevant to the NLO run.\\
$\blacklozenge$ 
{\variable FPREFIXxxx.data}: {\variable xxx} can assume several different 
values. These are the data files created by the \NLO\ code. They can be 
removed only if no further event generation step is foreseen with the
currect choice of parameters.\\
$\blacklozenge$ 
{\variable FPREFIXMCinput}: analogous to {\variable FPREFIXNLOinput}, 
but for the \MC\ executable. See table~\ref{tab:MCi}.\\
$\blacklozenge$ 
{\variable FPREFIXMC.log}: analogous to {\variable FPREFIXNLO.log}, but 
for the MC run.\\
$\blacklozenge$ 
{\variable EVPREFIX.events}: the event file, where {\variable EVPREFIX} 
is the string set by the user in {\code MCatNLO.inputs}.\\
$\blacklozenge$ 
{\variable EVPREFIXxxx.events}: {\variable xxx} can assume several different 
values. These files are temporary event files, which are used by the
NLO code, and eventually removed by the shell script. They MUST NOT be
removed by the user during the run (the program will crash or give
meaningless results).

By default, all the files produced by the
$\MCatNLO$ are written in the running directory.
However, if the variable {\variable SCRTCH} 
(to be set in {\code MCatNLO.inputs}) is {\em not} blank, 
the data and event files will be written in the directory 
whose address is stored in {\variable SCRTCH}.

\section{Script variables\label{sec:scrvar}}
In the following, we list all the variables appearing in 
{\code MCatNLO.inputs}; these can be changed by the user to suit 
his/her needs. This must be done by editing {\code MCatNLO.inputs}.
\begin{itemize}
\item[{\variable ECM}] 
 The CM energy of the colliding particles.
\item[{\variable FREN}] 
 The ratio between the renormalization scale, and a reference mass scale.
\item[{\variable FFACT}] 
 As {\variable FREN}, for the factorization scale.
\item[{\variable FRENMC}] 
 As {\variable FREN}; enters the MC-subtraction terms $\bSigma\xMC$
 (see ref~\cite{Frixione:2002ik}).
\item[{\variable FFACTMC}] 
 As {\variable FFACT}; enters the MC-subtraction terms $\bSigma\xMC$
 (see ref~\cite{Frixione:2002ik}).
\item[{\variable xMASS}] 
 The mass (in GeV) of the particle {\variable x}, with 
 {\variable x=W,Z,U,D,S,C,B,G}.
\item[{\variable IPROC}]
 Process number that identifies the vector bosons in the final 
 states: see table~\ref{tab:proc} for valid entries.
\item[{\variable PARTn}]
 The type of the incoming particle \#{\variable n}, with {\variable n}=1,2. 
 \HW\ naming conventions are used ({\variable P, PBAR, N, NBAR}).
\item[{\variable PDFGROUP}]
 The name of the group fitting the parton densities used;
 the labeling conventions of PDFLIB are adopted.
\item[{\variable PDFSET}] 
 The number of the parton density set; according to PFDLIB,
 the pair ({\variable PDFGROUP}, {\variable PDFSET}) identifies the densities.
\item[{\variable LAMBDAFIVE}]
 The value of $\Lambda_{\sss QCD}$, for five flavours and in the 
 ${\overline {\rm MS}}$ scheme.
\item[{\variable SCHEMEOFPDF}] 
 The subtraction scheme in which the parton densities are defined.
\item[{\variable FPREFIX}] Our integration routine creates files with
 names beginning with the string {\variable FPREFIX}. 
 Most of these files are not directly 
 accessed by the user; for more details, see sect.~\ref{sec:evfile}.
\item[{\variable EVPREFIX}] 
 The name of the event file begins with this string; for 
 more details, see sect.~\ref{sec:evfile}.
\item[{\variable NEVENTS}] 
 The number of events stored in the event file, eventually
 processed by \HW.
\item[{\variable WGTTYPE}]
 Valid entries are 0 and 1. When set to 0, the weights are $\pm 1$. When
 set to 1, the weights are $\pm w$, with $w$ a constant such that the sum 
 of the weights gives the total NLO rate.
\item[{\variable RNDEVSEED}] 
 This is the seed for the random number generation is the
 event generation step; must be changed in order to obtain
 statistically-equivalent but different event files.
\item[{\variable BASES}] 
 Controls the integration step; valid entries are {\variable ON} and 
 {\variable OFF}. At least one run with {\variable BASES=ON} must be 
 performed.
\item[{\variable PDFLIBRARY}] 
 Valid entries are {\variable PDFLIB} and {\variable THISLIB}. 
 In the former case,
 the local version of PDFLIB is used to compute the parton
 densities, whereas in the latter case the densities are
 obtained from our self-contained faster package.
\item[{\variable HERPDF}] 
 If set to {\variable DEFAULT}, \HW\ uses its internal PDF set 
 (controlled by {\variable NSTRU}), regardless of the densities
 adopted at the NLO level. If set to {\variable EXTPDF}, \HW\ uses
 the same PDFs as the NLO code.
\item[{\variable HWPATH}] 
 The physical address of the directory where the user's
 preferred version of \HW\ is stored.
\item[{\variable SCRTCH}]
 The physical address of the directory where the user wants to store the
 data and event files. If left blank, these files are stored in the 
 running directory.
\end{itemize}

\section{{\code Makefile} variables\label{sec:Makevar}}
Before running the package, the user must edit the {\code Makefile}, in order
to set the following variables. Further changes of the {\code Makefile} are 
necessary only if one wants to change the \HW\ version linked, or
other files must be added for new features in the analysis routines.
\begin{itemize}
\item[{\variable HWUTI}]
This variable must be set equal to a list of object files,
needed by the analysis routines of the user (for example,
{\variable HWUTI=obj1.o obj2.o obj3.o} is a valid assignment)
\item[{\variable HERWIGVER}]
This variable must to be set equal to the name of the
object file corresponding to the version of \HW\ linked
to the package (for example, {\variable HERWIGVER=herwig64.o} is a
valid assignment)
\end{itemize}

\appendix
\section{Running the package without the shell scripts\label{app:instr}}
In this appendix, we describe the actions that the user needs to 
take in order to run the package without using the shell scripts,
and the {\variable Makefile}.

\subsection{Creating the executables\label{app:exe}}
An $\MCatNLO$ run requires the creation of two executables, for the NLO
and MC codes respectively. The files to link depend on whether one
uses PDFLIB, or the PDF library provided with this package; we list
them below:
\begin{itemize}
\item {\bf NLO without PDFLIB:}
{\code mcatnlo\_vbmain.o mcatnlo\_vbxsec.o mcatnlo\_date.o mcatnlo\_int.o 
mcatnlo\_uxdate.o mcatnlo\_uti.o mcatnlo\_str.o\\
mcatnlo\_pdftomlm.o mcatnlo\_libofpdf.o dummies.o SYSFILE}
\item {\bf NLO with PDFLIB:}
{\code mcatnlo\_vbmain.o mcatnlo\_vbxsec.o mcatnlo\_date.o mcatnlo\_int.o 
mcatnlo\_uxdate.o mcatnlo\_uti.o mcatnlo\_str.o\\
mcatnlo\_mlmtopdf.o dummies.o}
{\variable SYSFILE CERNLIB}
\item {\bf MC without PDFLIB:}
{\code mcatnlo\_hwanal.o mcatnlo\_hwdriver.o mcatnlo\_hwhvvj.o 
mcatnlo\_str.o mcatnlo\_pdftomlm.o mcatnlo\_libofpdf.o dummies.o} 
{\variable HWUTI HERWIGVER}
\item {\bf MC with PDFLIB:}
{\code mcatnlo\_hwanal.o mcatnlo\_hwdriver.o mcatnlo\_hwhvvj.o 
mcatnlo\_str.o mcatnlo\_mlmtopdf.o dummies.o}
{\variable HWUTI HERWIGVER CERNLIB}
\end{itemize}
Here, {\variable SYSFILE} must be set either equal to {\code alpha.o},
or to {\code linux.o}, or to {\code sun.o}, according to the architecture 
of the machine on which the run is performed. For any other architecture,
the user should provide a file corresponding to {\code alpha.f} etc.,
which he/she will easily obtain by modifying {\code alpha.f}. The 
variables {\variable HWUTI} and {\variable HERWIGVER} have been described
in sect.~\ref{sec:Makevar}. Finally, the variable {\variable CERNLIB} must
be set in order to link the local version of CERN PDFLIB. To create
the object files eventually linked, static compilation is always
recommended (for example, {\code g77 -Wall -fno-automatic} on Linux).

\subsection{The input files\label{app:input}}
In this appendix, we describe the inputs to be given to the \NLO\ and 
\MC\ executables. When the shell scripts are used to run the $\MCatNLO$,
two files are created, {\variable FPREFIXNLOinput} and 
{\variable FPREFIXMCinput}, which are read by the \NLO\ and \MC\ executable
respectively. We start by considering the inputs for the \NLO\
executable, presented in table~\ref{tab:NLOi}.
\begin{table}[htb]
\begin{center}
\begin{tabular}{ll}
\hline
 '{\variable FPREFIX}'                       & ! prefix for BASES files\\
 '{\variable EVPREFIX}'                      & ! prefix for event files\\
  {\variable ECM FFACT FREN FFACTMC FRENMC}  & ! energy, scalefactors\\
  {\variable IPROC}                          & ! 2850/60/70/80=WW/ZZ/ZW+/ZW-\\
  {\variable WMASS ZMASS}                    & ! M\_W, M\_Z\\
  {\variable UMASS DMASS SMASS CMASS BMASS GMASS} & ! quark and gluon masses\\
 '{\variable PART1}'  '{\variable PART2}'    & ! hadron types\\
 '{\variable PDFGROUP}'   {\variable PDFSET} & ! PDF group and id number\\
  {\variable LAMBDAFIVE}                     & ! Lambda\_5, $<$0 for default\\
 '{\variable SCHEMEOFPDF}'                   & ! scheme\\
  {\variable NEVENTS}                        & ! number of events\\
  {\variable WGTTYPE}                      & ! 0 =$>$ wgt=+1/-1, 1 otherwise\\
  {\variable RNDEVSEED}                      & ! seed for rnd numbers\\
  {\variable zi}                             & ! zi\\
  {\variable nitn$_1$ nitn$_2$}              & ! itmx1,itmx2\\
\hline\\
\end{tabular}
\end{center}
\caption{\label{tab:NLOi}
Sample input file for the \NLO\ code. {\variable FPREFIX} and
{\variable EVPREFIX} must be understood with {\variable SCRTCH}
in front (see sect.~\ref{sec:scrvar}).
}
\end{table}
The variables whose name is in uppercase characters have been described 
in sect.~\ref{sec:scrvar}. The other variables are assigned by the shell
script. Their default values are given in table~\ref{tab:defNLO}.
\begin{table}[htb]
\begin{center}
\begin{tabular}{ll}
\hline
Variable & Default value\\
\hline
{\variable zi}          & 0.1\\
{\variable nitn$_i$}    & 10/0 ({\variable BASES=ON/OFF})\\
\hline\\
\end{tabular}
\end{center}
\caption{\label{tab:defNLO}
Default values for script-generated variables in {\code FPREFIXNLOinput}.
}
\end{table}
Users who run the package without the script should use the values
given in table~\ref{tab:defNLO}. The variable {\variable zi} controls,
to a certain extent, the number of negative-weight events generated 
by the $\MCatNLO$ (see ref.~\cite{Frixione:2002ik}). Therefore, the user
may want to tune this parameter in order to reduce as much as possible
the number of negative-weight events. We stress that the \MC\ code will
not change this number; thus, the tuning can (and must) be done only 
by running the \NLO\ code. The variables {\variable nitn$_i$} control
the integration step (see sect.~\ref{sec:evfile}), which can be
skipped by setting {\variable nitn$_i=0$}. If one needs to perform the
integration step, we suggest setting these variables as indicated in
table~\ref{tab:defNLO}. 

\begin{table}[htb]
\begin{center}
\begin{tabular}{ll}
\hline
 '{\variable EVPREFIX.events}'               & ! event file\\
  {\variable NEVENTS}                        & ! number of events\\
  {\variable esctype}         & ! 0-$>$EMSCA=sqrt(s)-2*pT, 1-$>$EMSCA=sqrt(s)\\
  {\variable pdftype}                      & ! 0-$>$Herwig PDFs, 1 otherwise\\
 '{\variable PART1}'  '{\variable PART2}'    & ! hadron types\\
  {\variable beammom beammom}                & ! beam momenta\\
  {\variable IPROC}                          & ! 2850/60/70/80=WW/ZZ/ZW+/ZW-\\
 '{\variable PDFGROUP}'                      & ! PDF group (1)\\
  {\variable PDFSET}                         & ! PDF id number (1)\\
 '{\variable PDFGROUP}'                      & ! PDF group (2)\\
  {\variable PDFSET}                         & ! PDF id number (2)\\
  {\variable LAMBDAFIVE}                     & ! Lambda\_5, $<$0 for default\\
  {\variable WMASS WMASS ZMASS}              & ! M\_W+, M\_W-, M\_Z\\
  {\variable UMASS DMASS SMASS CMASS BMASS GMASS} & ! quark and gluon masses\\
\hline\\
\end{tabular}
\end{center}
\caption{\label{tab:MCi}
Sample input file for the \MC\ code, resulting from setting 
{\variable HERPDF=DEFAULT}, which implies {\variable pdftype=1}. 
Setting {\variable HERPDF=EXTPDF} results in an analogous file, with
{\variable pdftype=0}, and without the lines concerning
{\variable PDFGROUP} and {\variable PDFSET}. {\variable EVPREFIX} 
must be understood with {\variable SCRTCH} in front 
(see sect.~\ref{sec:scrvar}).
}
\end{table}
We now turn to the inputs for the MC executable, presented
in table~\ref{tab:MCi}. 
The variables whose names are in uppercase characters have been described 
in sect.~\ref{sec:scrvar}. The other variables are assigned by the shell
script. Their default values are given in table~\ref{tab:defMC}.
\begin{table}[htb]
\begin{center}
\begin{tabular}{ll}
\hline
Variable & Default value\\
\hline
{\variable esctype}         & 0\\
{\variable pdftype}         & 0/1 ({\variable HERPDF=DEFAULT/EXTPDF})\\
{\variable beammom}         & {\variable EMC}/2\\
\hline\\
\end{tabular}
\end{center}
\caption{\label{tab:defMC}
Default values for script-generated variables in {\code MCinput}.
}
\end{table}
The user can freely change the values of {\variable esctype} and
{\variable pdftype}; on the other hand, the value of {\variable beammom}
must always be equal to half of the hadronic CM energy.

\end{document}